\documentclass[a4paper,12pt]{article}
\usepackage{epsfig}
\usepackage[dvips,usenames]{color}
\usepackage{graphicx}
\usepackage{color}
\usepackage{colortbl}

\newlength{\dinwidth}
\newlength{\dinmargin}
\newcommand{\spur}[1]{\not\! #1 \,}

\begin{document}
\title{Probe  R-parity Violating Supersymmetry Effects  in $B\to K^{(*)}\ell^+\ell^-$ and
$B_s\to \ell^+\ell^-$ Decays }

\author{Yuan-Guo Xu$^1$, Ru-Min
Wang$^{1,2}$  and Ya-Dong Yang$^1$\thanks{Corresponding author.
E-mail address: yangyd@henannu.edu.cn }
   \\
{\footnotesize {$^1$ \it Department of Physics, Henan Normal
University, XinXiang, Henan 453007, P.R.China}}
\\
 {\footnotesize {$^2$ \it Institute of Particle Physics,
 Huazhong Normal University,  Wuhan, Hubei 430070, P.R.China }}
  }

\maketitle \vspace{1.0cm}

\begin{abstract}
We study the decays $B\to K^{(*)}\ell^+\ell^-$ and $B_s\to
\ell^+\ell^-~(\ell=e,\mu)$ in the minimal supersymmetric standard
model with R-parity violation (RPV).  From the recent measurements
of their branching ratios,  we have derived new upper bounds on the
relevant RPV coupling products, which are stronger than the existing
ones. Using the constrained parameter space, we predict the RPV
effects on the forward-backward asymmetries  ${\mathcal A}_{FB}(B\to
K^{(*)}\ell^+\ell^- )$ and the branching ratios $ {\mathcal B}(
B_s\to \ell^{+}\ell^{-}) $. Our results of the forward-backward
asymmetries agree with the recent experiment data.  It is also found
that $\mathcal{B}(B_s\to \ell^+\ell^-)$ could be enhanced several
orders by the RPV  sneutrino  exchange.  The RPV effects on the
dilepton invariant mass spectra of $B\to K^{(*)}\ell^+\ell^-$ and
the normalized ${\mathcal A}_{FB}(B\to K^{(*)}\ell^+\ell^- )$ are
studied in detail.  Our results  could be used  to  probe RPV
effects and will correlate with searches  for direct RPV signals at
LHC.
\end{abstract}

\vspace{1cm} \noindent {\bf PACS Numbers 13.20.He  12.60.Jv 11.30.Er
12.15.Mm}

\newpage
\section{Introduction}
Flavor changing neutral current (FCNC) $b\to s$ processes
are forbidden at the tree level in the standard model (SM), which
proceed at a low rate via penguin or box diagrams. If additional
diagrams with non-SM particles contribute, their rates as well as other
properties will be modified. This feature make FCNC processes  powerful means
to probe new physics indirectly. The recent experimental
measurements of  $B\to K^{(*)}\ell^+\ell^-$ decays\cite{ba1,be1,be2,HFAG}
agree with the SM predictions within their
error bars, therefore, these measurements will afford an opportunity
to constrain new physics scenarios beyond the SM.

Semileptonic rare decays $B\to K^{(*)}\ell^+\ell^-$  have been extensively
studied previously. The dominant perturbative SM contribution had
been evaluated years ago\cite{Grinstein}, and later  QCD
corrections have been provided \cite{Buchalla,buraswilson,GAM}.
$\mathcal{O}(1/m^2_b)$ corrections have been first
calculated in Ref.\cite{Falk} and then in Refs.\cite{A. Ali and G.
Hiller,G. Buchalla and G. Isidori}.
 Long distance contributions  can have  different origins according to the value of the dilepton
invariant mass. The contributions of charmonium  resonances to these decays  by means of
Vector Meson Dominance (VMD) have been studied  carefully\cite{A. Ali and G.
Hiller, Deshpande, Kruger, att1991}.
Far from the resonance region, instead,
 $c\bar{c}$ long-distance effects  are investigated
using a heavy quark expansion in inverse powers of the charm-quark
mass ($\mathcal{O}(1/m^2_c)$ corrections) \cite{G.Buchalla}. Analyses  of  new physics
contribution have been performed  in different models, for example,
the two-Higgs doublet model\cite{dai} , the supersymmetric
(SUSY) models \cite{ea,ap}, the SUSY SO(10) grand unification theory
\cite{liwenjun} and the top quark two-Higgs doublet model
\cite{xiao}.

The effects  of RPV  SUSY in $B$ meson decays have been extensively
investigated in the literature \cite{RPVstudy,kundu}. The decays $B
\to K^{(*)}\ell^+\ell^-$ and $B_s\to \ell^+\ell^- $ are all induced
at the parton level by $b\to s\ell^+\ell^-$ process, and they
involve the same set of the RPV coupling products. In this paper we
will study the decays $B \to K^{(*)}\ell^+\ell^-$ and $B_s\to
\ell^+\ell^- $ in the RPV SUSY model. Using the recent experimental
data, we will obtain the new upper limits on the relevant RPV
coupling products. Then we will use the constrained regions of the
parameters to examine the RPV effects on the branching ratios of
$B_s \to \ell^+\ell^-$ decays and the forward-backward asymmetries
$(\mathcal{A}_{FB})$ of $B\to K^{(*)}\ell^+\ell^-$. In addition, we
will compare the SM predictions with the RPV predictions about
dilepton invariant mass spectra  and the normalized forward-backward
asymmetries in $B \to K^{(*)}\ell^+\ell^-$ decays.

The paper is arranged as follows. In Section 2,  we introduce the
effective Hamiltonian and calculate the expressions  for $B\to
K^{(*)}\ell^+\ell^-$ and $B_s\to \ell^+\ell^-$ processes in the RPV
SUSY.  In Section 3, we tabulate the theoretical inputs and deal with
the numerical results. We display the constrained parameter spaces
which satisfy all the available experimental data,  and then we use
the constrained  parameter spaces to  predict the RPV effects on
  $\mathcal{A}_{FB}$($B\to K^{(*)}\ell^+\ell^-$)
 and $\mathcal{B}(B_s\to \ell^+\ell^-)$, which have not been well measured
 yet. We also show the RPV effects on
dilepton invariant mass spectra  and the normalized forward-backward
asymmetries in $B \to K^{(*)}\ell^+\ell^-$ decays.
 Section 4 contains our summary and conclusion.

\section{The theoretical frame for $B \to K^{(*)}\ell^+\ell^-$ and $B_s\to \ell^+\ell^- $}
\subsection{The decay branching ratios in the SM}
\subsubsection{The semileptonic decays $B\to K^{(*)} \ell^+\ell^-$}

In the SM, at the quark level, the rare semileptonic decays $b\to s
\ell^+\ell^-$ can be described by  the effective Hamiltonian
\begin{eqnarray}
\mathcal{H}^{SM}_{eff}(b\to s
\ell^+\ell^-)=-\frac{G_F}{\sqrt{2}}V^*_{ts}V_{tb}\sum^{10}_{i=1}C_{i}(\mu)\mathcal{O}_{i},
\end{eqnarray}
where the operator base $\mathcal{O}_i$ is given in
\cite{buraswilson}. We will use  the Wilson coefficients $C_i(\mu)$
calculated in the naive dimensional regularization (NDR) scheme
\cite{buraswilson}, and
 the long-distance resonance effects on $C^{eff}_9(\mu)$  given in
 Refs.\cite{Deshpande,att1991}. The Hamiltonian leads to
the following free quark decay amplitude:
\begin{eqnarray}
\mathcal{M}^{SM}(b\to s
\ell^+\ell^-)&&=\frac{G_F\alpha_{e}}{\sqrt{2}\pi}V^*_{ts}V_{tb}\Big\{C^{eff}_9(\bar{s}\gamma_\mu
P_Lb)(\bar{\ell}\gamma^\mu \ell)+C_{10}(\bar{s}\gamma_\mu
P_Lb)(\bar{\ell}\gamma^\mu\gamma_5\ell)\nonumber\\
&&-2\hat{m}_bC^{eff}_7\left(\bar{s}i\sigma_{\mu\nu}\frac{\hat{q}^\nu}{\hat{s}}P_Rb\right)
(\bar{\ell}\gamma^\mu
\ell)\Big\},\label{quarkM}
\end{eqnarray}
with $ P_{L,R} \equiv (1\mp\gamma_5)/2, s = q^2$ and $q =
p_++p_-$ ($p_\pm$  the four-momenta of the leptons).  In our following calculations, we take
$m_s/m_b=0$, but keep the lepton masses. The hat denotes
normalization in terms of the B-meson mass, $m_B$, e.g.
$\hat{s}=s/m_B^2$, $\hat{m}_q=m_q/m_B$.

Exclusive decays $B\to K^{(*)}\ell^+\ell^-$ are described in terms
of matrix elements of the quark operators in Eq.(\ref{quarkM}) over
meson states, which can be parameterized  by the form factors. It is
worth noting the form factors involving the $B\to K^{(*)}$
transitions have been updated recently in \cite{BallZwicky}.

 Using Eq.(\ref{quarkM}), one can get the
amplitudes of exclusive $B\to K^{(*)}\ell^+\ell^-$ decays
\begin{eqnarray}
\mathcal{M}^{SM}(B\to
K^{(*)}\ell^+\ell^-)=\frac{G_F\alpha_{e}}{2\sqrt{2}~\pi}
V^*_{ts}V_{tb}m_B\left[\mathcal{T}_{1\mu}(\bar{\ell}\gamma^\mu
\ell)+\mathcal{T}_{2\mu}(\bar{\ell}\gamma^\mu \gamma_5\ell)\right],
\end{eqnarray}
where for $B\to K\ell^+\ell^-$,
\begin{eqnarray}
\mathcal{T}_{1\mu}=A'(\hat{s})\hat{p}_\mu+B'(\hat{s})\hat{q}_\mu,\\
\mathcal{T}_{2\mu}=C'(\hat{s})\hat{p}_\mu+D'(\hat{s})\hat{q}_\mu,
\end{eqnarray}
and for $B \to K^*\ell^+\ell^-$,
\begin{eqnarray}
\mathcal{T}_{1\mu}&=&A(\hat{s})\epsilon_{\mu\rho\alpha\beta}\epsilon^{*\rho}
\hat{p}^\alpha_B\hat{p}^\beta_{K^*}-iB(\hat{s})\epsilon^*_\mu
+iC(\hat{s})(\epsilon^*\cdot\hat{p}_B)\hat{p}_\mu+iD(\hat{s})(\epsilon^*\cdot\hat{p}_B)\hat{q}_\mu,\\
\mathcal{T}_{2\mu}&=&E(\hat{s})\epsilon_{\mu\rho\alpha\beta}
\epsilon^{*\rho}\hat{p}^\alpha_B\hat{p}^\beta_{K^*}-iF(\hat{s})\epsilon^*_\mu
+iG(\hat{s})(\epsilon^*\cdot\hat{p}_B)\hat{p}_\mu+iH(\hat{s})(\epsilon^*\cdot\hat{p}_B)\hat{q}_\mu,
\end{eqnarray}
with $ p = p_B+p_{K^{(*)}}$. Note that, using the equation of
motion for lepton fields, the  $\hat{q}_\mu$ terms  in
$\mathcal{T}_{1\mu}$  vanish,  and those in $\mathcal{T}_{2\mu}$
become suppressed by one power of the lepton mass.

 The auxiliary functions in $\mathcal{T}_{1\mu}$ and $\mathcal{T}_{2\mu}$ are defined
 as \cite{ap}
\begin{eqnarray}
A'(\hat{s})&=&C^{eff}_9(\hat{s})f_+(\hat{s})
+\frac{2\hat{m}_b}{1+\hat{m}_K}C^{eff}_7f_T(\hat{s}),\\
B'(\hat{s})&=&C^{eff}_9(\hat{s})f_-(\hat{s})
-\frac{2\hat{m}_b}{\hat{s}}(1-\hat{m}_K)C^{eff}_7f_T(\hat{s}),\\
C'(\hat{s})&=&C_{10}f_+(\hat{s}),\\
D'(\hat{s})&=&C_{10}f_-(\hat{s}),\\
A(\hat{s})&=&\frac{2}{1+\hat{m}_{K^*}}C^{eff}_9(\hat{s})
V(\hat{s})+\frac{4\hat{m}_b}{\hat{s}}C^{eff}_7T_1(\hat{s}),\\
B(\hat{s})&=&(1+\hat{m}_{K^*})\left[C^{eff}_9(\hat{s})A_1(\hat{s})
+\frac{2\hat{m}_b}{\hat{s}}(1-\hat{m}_{K^*})C^{eff}_7T_2(\hat{s})\right],\\
C(\hat{s})&=&\frac{1}{1-\hat{m}^2_{K^*}}\left[(1-\hat{m}_{K^*})C^{eff}_9(\hat{s})A_2(\hat{s})
+2\hat{m}_bC^{eff}_7\left(T_3(\hat{s})
+\frac{1-\hat{m}_{K^*}}{\hat{s}}T_2(\hat{s})\right)\right],\\
D(\hat{s})&=&\frac{1}{\hat{s}}\Big[C^{eff}_{9}(\hat{s})
\Big((1+\hat{m}_{K^*})A_1(\hat{s})-(1-\hat{m}_{K^*})A_2(\hat{s})
-2\hat{m}_{K^*}A_0(\hat{s})\Big)\nonumber\\&&-2\hat{m}_bC^{eff}_7T_3(\hat{s})\Big],\\
E(\hat{s})&=&\frac{2}{1+\hat{m}_{K^*}}C_{10}V(\hat{s}),\\
F(\hat{s})&=&(1+\hat{m}_{K^*})C_{10}A_1(\hat{s}),\\
G(\hat{s})&=&\frac{1}{1+\hat{m}_{K^*}}C_{10}A_2(\hat{s}),\\
H(\hat{s})&=&\frac{1}{\hat{s}}C_{10}\Big[(1+\hat{m}_{K^*})A_1(\hat{s})
-(1-\hat{m}_{K^*})A_2(\hat{s})-2\hat{m}_{K^*}A_0(\hat{s})\Big].
\end{eqnarray}
It's noted that the inclusion of the full s-quark mass dependence in
the above formulae can be done by substituting $m_b\to
m_b+m_s $ into all terms proportional to $C_7^{eff}T_1$ and
$C^{eff}_7f_T$ and $ m_b\to m_b-m_s $ in $C_7^{eff}T_{2,3}$,
since $\mathcal{O}_7\sim
\overline{s}\sigma_{\mu\nu}[(m_b+m_s)+(m_b-m_s)\gamma_5]q^{\nu}b$.

The kinematic variables $(\hat{s},\hat{u})$ are chosen to be
\begin{eqnarray}
\hat{s}&=&\hat{q}^2=(\hat{p}_++\hat{p}_-)^2,\\
\hat{u}&=&(\hat{p}_B-\hat{p}_-)^2-(\hat{p}_B-\hat{p}_+)^2,
\end{eqnarray}
which are bounded as
\begin{eqnarray}
(2\hat{m}_\ell)^2\leq&\hat{s}&\leq(1-\hat{m}_{K^{(*)}})^2,\\
-\hat{u}(\hat{s})\leq&\hat{u}&\leq\hat{u}(\hat{s}),
\end{eqnarray}
with $\hat{m}_\ell=m_{\ell}/m_B$ and
\begin{eqnarray}
\hat{u}(\hat{s})&=&\sqrt{\lambda\big(1-4\frac{\hat{m}^2_{\ell}}{\hat{s}}\big)},\\
\lambda&\equiv &\lambda(1,\hat{m}^2_{K^{(*)}},\hat{s})\nonumber\\
&=&1+\hat{m}^4_{K^{(*)}}
+\hat{s}^2-2\hat{s}-2\hat{m}^2_{K^{(*)}}(1+\hat{s}).
\end{eqnarray}
The variable $\hat{u}$ corresponds to $\theta$, the angle between
the momentum of the B-meson and the  lepton
$\ell^+$ in the dilepton center-of-mass system (CMS) frame,
through the relation $\hat{u} =-\hat{u}(s)\mbox{cos}\theta $
\cite{att1991}. Keeping the lepton mass, we find the double
differential decay branching ratios $\mathcal{B}^K$ and
$\mathcal{B}^{K^*}$ for the decays $B\to K\ell^+\ell^-$
and $B\to K^*\ell^+\ell^-$, respectively, as
\begin{eqnarray}
\frac{d^2\mathcal{B}^{K}_{SM}}{d\hat{s}d\hat{u}}&=&\tau_B
\frac{G^2_F\alpha_{e}^2m_B^5}{2^{11}\pi^5}|V^*_{ts}V_{tb}|^2 \nonumber\\
&&\times\Bigg\{(|A'|^2+|C'|^2)(\lambda-\hat{u}^2)\nonumber\\
&&+|C'|^24\hat{m}^2_\ell(2+2\hat{m}^2_K-\hat{s})
+Re(C'D'^*)8\hat{m}^2_\ell(1-\hat{m}^2_K)+|D'|^24\hat{m}^2_\ell\hat{s}\Bigg\},\\
\frac{d^2\mathcal{B}^{K^*}_{SM}}{d\hat{s}d\hat{u}}
&=&\tau_B\frac{G^2_F\alpha_{e}^2m_B^5}{2^{11}\pi^5}|V^*_{ts}V_{tb}|^2\nonumber\\
&&\times\left\{\frac{|A|^2}{4}\Big(\hat{s}(\lambda+\hat{u}^2)+4\hat{m}^2_\ell\lambda\Big)
+\frac{|E|^2}{4}\Big(\hat{s}(\lambda+\hat{u}^2)
-4\hat{m}^2_\ell\lambda\Big)\right.\nonumber\\
&&+\frac{1}{4\hat{m}^2_{K^*}}\Big[|B|^2\Big(\lambda-\hat{u}^2
+8\hat{m}^2_{K^*}(\hat{s}+2\hat{m}^2_\ell)\Big)
+|F|^2\Big(\lambda-\hat{u}^2+8\hat{m}^2_{K^*}(\hat{s}-4\hat{m}^2_\ell)\Big)\Big]\nonumber\\
&&-2\hat{s}\hat{u}\Big[Re(BE^*)+Re(AF^*)\Big]\nonumber\\
&&+\frac{\lambda}{4\hat{m}^2_{K^*}}\Big[|C|^2(\lambda-\hat{u}^2)+
|G|^2(\lambda-\hat{u}^2+4\hat{m}^2_\ell(2+2\hat{m}^2_{K^*}-\hat{s})\Big)\Big]\nonumber\\
&&-\frac{1}{2\hat{m}^2_{K^*}}\Big[Re(BC^*)(1-\hat{m}^2_{K^*}-\hat{s})(\lambda-\hat{u}^2)\nonumber\\
&&~~~~~~~~~+Re(FG^*)\Big((1-\hat{m}^2_{K^*}
-\hat{s})(\lambda-\hat{u}^2)+4\hat{m}^2_\ell\lambda\Big)\Big]\nonumber\\
&&\left.-2\frac{\hat{m}^2_\ell}{\hat{m}^2_{K^*}}
\lambda\Big[Re(FH^*)-Re(GH^*)(1-\hat{m}^2_{K^*})\Big]
+|H|^2\frac{\hat{m}^2_\ell}{\hat{m}^2_{K^*}}\hat{s}\lambda
\right\}.\label{BK}
\end{eqnarray}
Our results of the double differential decay branching ratios are
consistent with the ones in Ref.\cite{ap}.

\subsubsection{The pure leptonic decays $B_s\to \ell^+\ell^-$ }

In the SM, the effective Hamiltonian have been given by \cite{gbab}
\begin{eqnarray}
 \mathcal{H}^{SM}_{eff}(B_s\to \ell^+\ell^-)
 =-\frac{G_{F}}{\sqrt{2}}\frac{\alpha_{e}}{2 \pi \mbox{sin}^{2}\theta_W}V^{*}_{ts}
 V_{tb}Y(x_t)(\overline{s}b)_{V-A}(\overline{\ell}\ell)_{V-A}+h.c.,\label{pureH}
\end{eqnarray}
where $x_t=\frac{m_t^2}{m_W^2},m_t\equiv \overline{m}_t(m_t)$,
$(\bar{s}b)_{V-A}\equiv\bar{s}\gamma_{\mu}(1-\gamma_5)b$.
The pure leptonic decay amplitudes can be written as
 \begin{eqnarray}
\mathcal{M}^{SM}(B_s\to
\ell^+\ell^-)=h_{SM}\left(\bar{\ell}\spur{p}_{B}(1-\gamma_5)\ell\right),
\end{eqnarray}
with
\begin{eqnarray}
h_{SM}=-\frac{G_F}{\sqrt{2}}\frac{\alpha_{e}}{2\pi\mbox{sin}^2\theta_W}V^{*}_{ts}
 V_{tb}Y(x_t)(if_{B_s}).
\end{eqnarray}
Then we can get the branching ratios for $B_s\to\ell^+\ell^-$
\begin{eqnarray}
\mathcal{B}^{SM}(B_s\to \ell^+\ell^-)&=&\frac{\tau_{B_s}}{16\pi
m_{B_s}}\sqrt{1-4\hat{m}^2_\ell}\left|\mathcal{M}^{SM}(B_s\to
\ell^+\ell^-)\right|^2\nonumber\\
&=&\tau_{B_s}\frac{G^2_F}{\pi}\left(\frac{\alpha_{e}}{4\pi
\mbox{sin}^2\theta_W}\right)^2f^2_{B_s}m^2_{\ell}m_{B_s}
\sqrt{1-4\hat{m}^2_{\ell}}\left|V^*_{ts}V_{tb}\right|^2Y^2(x_t).
\end{eqnarray}

\subsection{The decay amplitudes in  RPV SUSY}
In the most general superpotential of the minimal supersymmetric
Standard Model (MSSM), the RPV superpotential is given by
\cite{RPVSW}
\begin{eqnarray}
\mathcal{W}_{\spur{R_p}}&=&\mu_i\hat{L}_i\hat{H}_u+\frac{1}{2}
\lambda_{[ij]k}\hat{L}_i\hat{L}_j\hat{E}^c_k+
\lambda'_{ijk}\hat{L}_i\hat{Q}_j\hat{D}^c_k+\frac{1}{2}
\lambda''_{i[jk]}\hat{U}^c_i\hat{D}^c_j\hat{D}^c_k, \label{rpv}
\end{eqnarray}
where $\hat{L}$ and $\hat{Q}$ are the SU(2) doublet lepton and quark
superfields, respectively. $\hat{E}^c$, $\hat{U}^c$ and $\hat{D}^c$
are the singlet superfields, while $i$, $j$ and $k$ are generation
indices and $c$ denotes a charge conjugate field.

The bilinear RPV superpotential terms $\mu_i\hat{L}_i\hat{H}_u$
can be rotated away by suitable redefining the lepton and Higgs
superfields \cite{barbier}. However, the rotation will generate a
soft SUSY breaking bilinear term which would affect our
calculation through loop level. However, the processes
discussed in this paper could be induced by tree-level RPV
couplings, so that we would neglect sub-leading RPV loop
contributions in this study.

The $\lambda$ and $\lambda'$ couplings in Eq.(\ref{rpv}) break the
lepton number, while the $\lambda''$ couplings break the baryon
number. There are 27 $\lambda'_{ijk}$ couplings, 9 $\lambda_{ijk}$
and 9 $\lambda''_{ijk}$ couplings.  $\lambda_{[ij]k}$ are
antisymmetric with respect to their first two indices, and
$\lambda''_{i[jk]}$ are antisymmetric with $j$ and $k$.

\begin{figure}[ht]
\begin{center}
\begin{tabular}{c}
\includegraphics[scale=1]{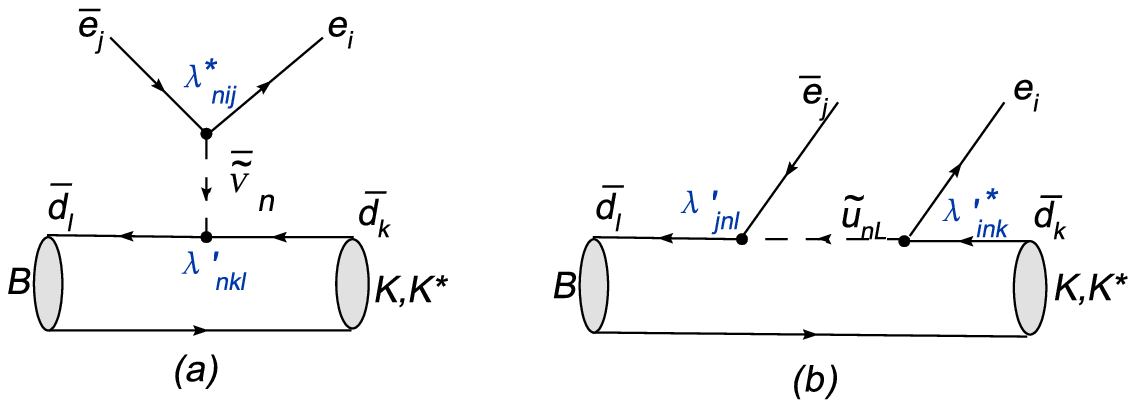}
\end{tabular}
\end{center}
\vspace{-0.8cm} \caption{ The RPV contributions to $B\to
K^{(*)}\ell^+\ell^-$ due to  sneutrino  and squark exchange.}
\label{semifig}
\begin{center}
\begin{tabular}{c}
\includegraphics[scale=1]{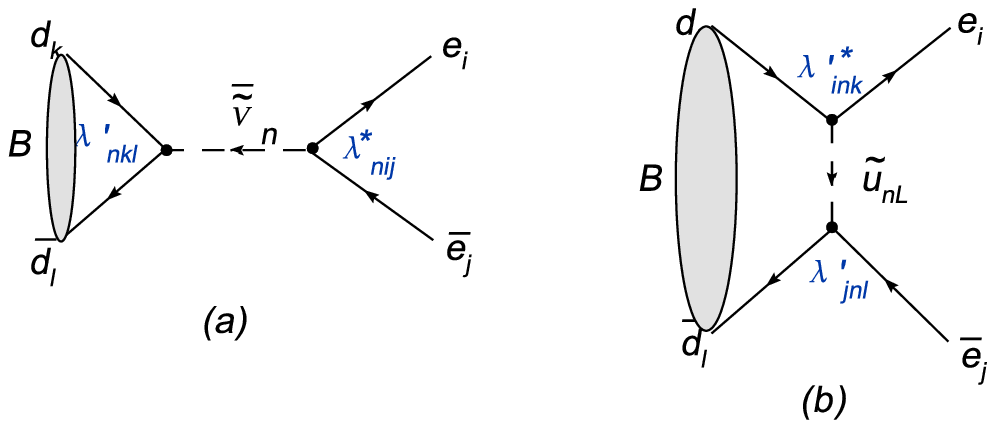}
\end{tabular}
\end{center}
\vspace{-0.8cm} \caption{ The RPV contributions to $B_s\to
\ell^+\ell^-$ due to  sneutrino  and squark exchange.}
\label{purefig}
\end{figure}
From Eq.(\ref{rpv}), we can obtain the relevant four fermion
effective Hamiltonian for  $b\to s\ell\ell$ process due to the
squarks and sneutrinos exchanges
\begin{eqnarray}
 \mathcal{H}^{\spur{R_p}}_{eff}&=&\frac{1}{2}\sum_i \frac{\lambda'_{jik}\lambda'^{*}_{lin}}
 {m^2_{\tilde{u}_{iL}}}(\bar{d}_k\gamma^\mu P_R d_n)(\bar{\ell}_l\gamma_\mu
P_L\ell_j) \nonumber\\
 && +\frac{1}{2}\sum_i\left\{\frac{\lambda_{ijk}\lambda'^{*}_{imn}}{m^2_{\tilde{\nu}_{iL}}}
 (\bar{d}_mP_Rd_n)(\bar{\ell}_kP_L\ell_j)+
\frac{\lambda^*_{ijk}\lambda'_{imn}}{m^2_{\tilde{\nu}_{iL}}}
(\bar{d}_nP_Ld_m)(\bar{\ell}_jP_R\ell_k)\right\}.
\label{RPVH}
\end{eqnarray}

 The RPV feynman diagrams for  $B\to K^{(*)}\ell^+\ell^-$ and
  $B_s\to \ell^+\ell^-$ are displayed in Fig.\ref{semifig} and
 Fig.\ref{purefig}, respectively.

From Eq.(\ref{RPVH}), we can obtain the RPV decay amplitude for
$B\to K\ell^+\ell^-$
\begin{eqnarray}
\mathcal{M}^{\spur{R_p}}(B\to
K\ell^+\ell^-)=\Delta_{\tilde{u}}\Big(\bar{\ell}_k(\spur{p}_B+\spur{p}_K)(1-\gamma_5)\ell_j\Big)
+\Delta_{\tilde{\nu}}\Big(\bar{\ell}_k(1-\gamma_5)\ell_j\Big)
+\Delta'_{\tilde{\nu}}\Big(\bar{\ell}_k(1+\gamma_5)\ell_j\Big),
\end{eqnarray}
with
\begin{eqnarray}
&&\Delta_{\tilde{u}}=\sum_i\frac{\lambda'_{ji3}\lambda_{ki2}'^{*}}{8m^2_{\tilde{u}_{iL}}}
f_+^{B\to K}(\hat{s}),\\
&&\Delta_{\tilde{\nu}}=\sum_i\frac{\lambda_{ijk}\lambda_{i32}'^{*}}{8m^2_{\tilde{\nu}_{iL}}}
f_+^{B\to K}(\hat{s})\frac{m^2_B-m^2_K}{\overline{m}_b-\overline{m}_s},\\
&&\Delta'_{\tilde{\nu}}=\sum_i\frac{\lambda^*_{ikj}\lambda'_{i23}}{8m^2_{\tilde{\nu}_{iL}}}
f_+^{B\to
K}(\hat{s})\frac{m^2_B-m^2_K}{\overline{m}_b-\overline{m}_s},
\end{eqnarray}
and  $\overline{m}_b$ and $\overline{m}_s$  the quark's running
masses at the scale $m_b$.

 For $B\to K^*\ell^+\ell^-$, the RPV amplitude is
\begin{eqnarray}
\mathcal{M}^{\spur{R_p}}(B\to
K^*\ell^+\ell^-)=\mathcal{T}_{3\mu}\Big(\bar{\ell}_k\gamma^\mu(1-\gamma_5)\ell_j\Big)
+~\Omega_{\tilde{\nu}}\Big(\bar{\ell}_k(1-\gamma_5)\ell_j\Big)
+~\Omega'_{\tilde{\nu}}\Big(\bar{\ell}_k(1+\gamma_5)\ell_j\Big),
\end{eqnarray}
where
\begin{eqnarray}
\mathcal{T}_{3\mu}&=&I(\hat{s})\epsilon_{\mu\rho\alpha\beta}\epsilon^{*\rho}
\hat{p}^\alpha_B\hat{p}^\beta_{K^*}-iJ(\hat{s})\epsilon^*_\mu
+iK(\hat{s})(\epsilon^*\cdot\hat{p}_B)\hat{p}_\mu+iL(\hat{s})(\epsilon^*\cdot\hat{p}_B)\hat{q}_\mu,\\
\Omega_{\tilde{\nu}}&=&\sum_i\frac{\lambda_{ijk}\lambda_{i32}'^{*}}{8m^2_{\tilde{\nu}_{iL}}}
\left[-\frac{i}{2}\frac{A_0^{B\to K^*}(\hat{s})}{\overline{m}_b+\overline{m}_s}
\lambda^{\frac{1}{2}}m^2_B\right],\\
\Omega'_{\tilde{\nu}}&=&\sum_i\frac{\lambda^*_{ikj}\lambda'_{i23}}{8m^2_{\tilde{\nu}_{iL}}}
\left[\frac{i}{2}\frac{A_0^{B\to
K^*}(\hat{s})}{\overline{m}_b+\overline{m}_s}\lambda^{\frac{1}{2}}m^2_B\right],
\end{eqnarray}
and the auxiliary functions are given as
\begin{eqnarray}
I(\hat{s})&=&\sum_i\frac{\lambda'_{ji3}\lambda_{ki2}'^{*}}{8m^2_{\tilde{u}_{iL}}}
\left[\frac{2V^{B\to K^*}(\hat{s})}{m_B+m_{K^*}}m^2_B\right],\\
J(\hat{s})&=&\sum_i\frac{\lambda'_{ji3}\lambda_{ki2}'^{*}}{8m^2_{\tilde{u}_{iL}}}
\left[-(m_B+m_{K^*})A_1^{B\to K^*}(\hat{s})\right],\\
K(\hat{s})&=&\sum_i\frac{\lambda'_{ji3}\lambda_{ki2}'^{*}}{8m^2_{\tilde{u}_{iL}}}
\left[\frac{A_2^{B\to K^*}(\hat{s})}{m_B+m_{K^*}}m_B^2\right],\\
L(\hat{s})&=&\sum_i\frac{\lambda'_{ji3}\lambda_{ki2}'^{*}}{8m^2_{\tilde{u}_{iL}}}
\left[\frac{~2m_{K^*}}{\hat{s}}\Big(A_3^{B\to
K^*}(\hat{s})-A_0^{B\to K^*}(\hat{s})\Big)\right].
\end{eqnarray}

For  $B_s\to \ell^+\ell^-$, the RPV amplitude is
\begin{eqnarray}
\mathcal{M}^{\spur{R_p}}(B_s\to
\ell^+\ell^-)=\Lambda_{\tilde{u}}\Big(\bar{\ell}_k\spur{p}_B(1-\gamma_5)\ell_j\Big)
+\Lambda_{\tilde{\nu}}\Big(\bar{\ell}_k(1-\gamma_5)\ell_j\Big)
+\Lambda'_{\tilde{\nu}}\Big(\bar{\ell}_k(1+\gamma_5)\ell_j\Big),
\end{eqnarray}
with
\begin{eqnarray}
\Lambda_{\tilde{u}}&=&\sum_i\frac{\lambda'_{ji3}\lambda_{ki2}'^{*}}{8m^2_{\tilde{u}_{iL}}}(-if_{B_s}),\\
\Lambda_{\tilde{\nu}}&=&\sum_i\frac{\lambda_{ijk}\lambda_{i32}'^{*}}{8m^2_{\tilde{\nu}_{iL}}}(-if_{B_s}\mu_{B_s}),\\
\Lambda'_{\tilde{\nu}}&=&\sum_i\frac{\lambda^*_{ikj}\lambda'_{i23}}{8m^2_{\tilde{\nu}_{iL}}}(if_{B_s}\mu_{B_s}),
\end{eqnarray}
and  $\mu_{B_s}\equiv\frac{m_{B_s}^2}{\overline{m}_b+\overline{m}_s}$.

 The RPV couplings can  be
complex in general,   we write their products as
\begin{eqnarray}
\Lambda_{ijk}\Lambda^*_{lmn} = |\Lambda_{ijk}\Lambda^*_{lmn}|~e^{i
\phi_{\spur{R_p}}},~~~~~\Lambda^*_{ijk}\Lambda_{lmn} =
|\Lambda_{ijk}\Lambda_{lmn}^*|~e^{-i \phi_{\spur{R_p}}},
\end{eqnarray}
where  the RPV coupling constant $\Lambda \in \{\lambda, \lambda'
\}$, and $\phi_{\spur{R_p}}$ is the RPV weak phase.

\subsection{The branching ratios with RPV contributions}

 With these formulae in Sec.2.1-2.2, we can obtain the total
double differential decay branching ratios of the decays $B\to
K^{(*)}\ell^+\ell^-$,
\begin{eqnarray}
\frac{d^2\mathcal{B}^{K^{(*)}}_{All}}{d\hat{s}d\hat{u}}
=\frac{d^2\mathcal{B}^{K^{(*)}}_{SM}}{d\hat{s}d\hat{u}}
+\frac{d^2\mathcal{B}^{K^{(*)}}_{\tilde{u}}}{d\hat{s}d\hat{u}}+
\frac{d^2\mathcal{B}^{K^{(*)}}_{\tilde{\nu}}}{d\hat{s}d\hat{u}}+
\frac{d^2\mathcal{B}'^{K^{(*)}}_{\tilde{\nu}}}{d\hat{s}d\hat{u}}.
\end{eqnarray}
Since we will only consider one RPV coupling product contributes at
one time, we have neglected the interferences between different RPV
coupling products, but kept their interferences  with the SM
amplitude, as shown in the following equations.

For the $B\to K\ell^+\ell^-$ decay,
\begin{eqnarray}
\frac{d^2\mathcal{B}^{K}_{\tilde{u}}}{d\hat{s}d\hat{u}}
&=&\tau_B\frac{m^4_B}{2^7\pi^3}\left\{\frac{}{}\right.
Re(WA'\Delta^*_{\tilde{u}})(\lambda-\hat{u}^2)\nonumber\\
&&+Re(WC'\Delta^*_{\tilde{u}}
)\Big[-(\lambda-\hat{u}^2)-4\hat{m}^2_{\ell}
(2+2\hat{m}^2_K-\hat{s})\Big]\nonumber\\
&&+Re(WD'\Delta^*_{\tilde{u}}
)\Big[-4\hat{m}^2_{\ell}(1-\hat{m}^2_K)\Big]\nonumber\\
&&+|\Delta_{\tilde{u}}|^2m_B\Big[\lambda-\hat{u}
+2\hat{m}^2_{\ell}(2+2\hat{m}^2_K-\hat{s})\Big]\left.\frac{}{}\right\},
\label{RPVBKmu}\\
\frac{d^2\mathcal{B}^{K}_{\tilde{\nu}}}{d\hat{s}d\hat{u}}
&=&\tau_B\frac{m^3_B}{2^7\pi^3}\left\{\frac{}{}\right.
Re(WA'\Delta^*_{\tilde{\nu}})(2\hat{m}_{\ell}\hat{u})
+Re(WC'\Delta^*_{\tilde{\nu}}
)(1-\hat{m}_K^2)(-2\hat{m}_{\ell})\nonumber\\
&&+Re(WD'\Delta^*_{\tilde{\nu}})(-2\hat{m}_{\ell}\hat{s})
+|\Delta_{\tilde{\nu}}|^2(\hat{s}-2\hat{m}^2_{\ell})\left.\frac{}{}\right\},\\
\frac{d^2\mathcal{B}^{K}_{\tilde{\nu}}}{d\hat{s}d\hat{u}}
&=&\tau_B\frac{m^3_B}{2^7\pi^3}\left\{\frac{}{}\right.
Re(WA'\Delta^*_{\tilde{\nu}})(2\hat{m}_{\ell}\hat{u})
+Re(WC'\Delta^*_{\tilde{\nu}}
)(1-\hat{m}_K^2)(2\hat{m}_{\ell})\nonumber\\
&&+Re(WD'\Delta^*_{\tilde{\nu}} )(2\hat{m}_{\ell}\hat{s})
+|\Delta_{\tilde{\nu}}|^2(\hat{s}-2\hat{m}^2_{\ell})\left.\frac{}{}\right\},
\end{eqnarray}
with $W=-\frac{G_F\alpha_{e}}{2\sqrt{2}~\pi}V^*_{ts}V_{tb}m_B$.

 For the $B\to K^*\ell^+\ell^-$ decay, we have
\begin{eqnarray}
\frac{d^2\mathcal{B}^{K^*}_{\tilde{u}}}{d\hat{s}d\hat{u}}&=&
\tau_B\frac{m_B^3}{2^9\pi^3}  \Bigg\{
Re(WAI^*)
\Big[\hat{s}(\lambda+\hat{u}^2)+4\hat{m}^2_{\ell}\lambda\Big]\nonumber\\
&&-Re(WEI^*)\Big[\hat{s}(\lambda+\hat{u}^2)-4\hat{m}^2_{\ell}\lambda\Big]+|I|^2
\Big[\hat{s}(\lambda+\hat{u}^2)\Big]\nonumber\\
&&+4\hat{s}\hat{u}\Big[Re(WAJ^*)+Re(WBI^*)-Re(WEJ^*)-Re(WFI^*)+2Re
(IJ^*)\Big]\nonumber\\
&&+\frac{1}{\hat{m}^2_{K^*}}\Bigg[Re(WBJ^*)\Big(\lambda-\hat{u}^2
+8\hat{m}^2_{K^*}(\hat{s}+2m_{\ell}^2)\Big)\nonumber\\
&&-Re(WFJ^*)\Big(\lambda-\hat{u}^2+8\hat{m}^2_{K^*}(\hat{s}-4m_{\ell}^2)\Big)\nonumber\\
&&+|J|^2\Big(\lambda-\hat{u}^2+8\hat{m}^2_{K^*}(\hat{s}-\hat{m}_{\ell}^2)\Big)\nonumber\\
&&-Re(WBK^*)(\lambda-\hat{u}^2)(1-\hat{m}^2_{K^*}-\hat{s})\nonumber\\
&&+Re(WFK^*)\Big((\lambda-\hat{u}^2)(1-\hat{m}^2_{K^*}-\hat{s})+4\hat{m}^2_{\ell}\lambda\Big)\nonumber\\
&&-2Re(JK^*)\Big((\lambda-\hat{u}^2)(1-\hat{m}^2_{K^*}-\hat{s})+2\hat{m}^2_{\ell}\lambda\Big)\Bigg]\nonumber\\
&&+\frac{\lambda}{\hat{m}^2_{K^*}}\Bigg[Re(WCK^*)(\lambda-\hat{u}^2)-Re(WGK^*)
\Big(\lambda-\hat{u}^2+4\hat{m}^2_{\ell}(2+2\hat{m}^2_{K^*}-\hat{s})\Big)\nonumber\\
&&+|K|^2\Big(\lambda-\hat{u}^2+2\hat{m}^2_{\ell}(2+2\hat{m}^2_{K^*}-\hat{s})\Big)\Bigg]\nonumber\\
&&+\frac{4\hat{m}^2_{\ell}}{\hat{m}^2_{K^*}}\lambda\Bigg[-Re(WHL^*)\hat{s}+|L|^2\hat{s}/2
+Re(WFL^*)-Re(JL^*)\nonumber\\
&&-Re(WGL^*)(1-\hat{m}^2_{K^*})+Re(KL^*)(1-\hat{m}^2_{K^*})\Bigg] \Bigg\},\\
\frac{d^2\mathcal{B}^{K^*}_{\tilde{\nu}}}{d\hat{s}d\hat{u}}&=&\tau_B\frac{m^3_B}{2^7\pi^3}
\Bigg\{-\frac{\hat{m}^2_{\ell}}{\hat{m}^2_{K^*}}
\Bigg[Im(WB\Omega_{\tilde{\nu}}^*)\cdot
\Big(\lambda^{-\frac{1}{2}}\hat{u}(1-\hat{m}^2_{K^*}-\hat{s})\Big)\nonumber\\
&&+Im(WC\Omega_{\tilde{\nu}}^*)\lambda^{\frac{1}{2}}\hat{u}
+Im(WF\Omega_{\tilde{\nu}}^*)\lambda^{\frac{1}{2}}\nonumber\\
&&-Im(WG\Omega_{\tilde{\nu}}^*)\lambda^{\frac{1}{2}}(1-\hat{m}^2_{K^*})\Bigg]
+|\Omega_{\tilde{\nu}}|^2(\hat{s}-2\hat{m}^2_{\ell})\Bigg\},\\
\frac{d^2\mathcal{B}'^{K^*}_{\tilde{\nu}}}{d\hat{s}d\hat{u}}&=&\tau_B\frac{m^3_B}{2^7\pi^3}
\Bigg\{-\frac{\hat{m}^2_{\ell}}{\hat{m}^2_{K^*}}
\Bigg[Im(WB\Omega'^*_{\tilde{\nu}})\cdot
\Big(\lambda^{-\frac{1}{2}}\hat{u}(1-\hat{m}^2_{K^*}-\hat{s})\Big)\nonumber\\
&&+Im(WC\Omega'^*_{\tilde{\nu}})
\lambda^{\frac{1}{2}}\hat{u}-Im(WF\Omega'^*_{\tilde{\nu}})\lambda^{\frac{1}{2}}\nonumber\\
&&+Im(WG\Omega'^*_{\tilde{\nu}})\lambda^{\frac{1}{2}}(1-\hat{m}^2_{K^*})\Bigg]
+|\Omega'_{\tilde{\nu}}|^2(\hat{s}-2\hat{m}^2_{\ell})\Bigg\}.
\end{eqnarray}

From  the total double differential branching ratios,  we can get the normalized
forward-backward asymmetries $\mathcal{A}_{FB}$ \cite{ba1}
\begin{eqnarray}
\mathcal{A}_{FB}(B\to K^{(*)}\ell^+\ell^-)=\int
d\hat{s}~\frac{\int^{+1}_{-1}\frac{d^2\mathcal{B}(B\to
K^{(*)}\ell^+\ell^-)}{d\hat{s}dcos\theta}sign(cos\theta)dcos\theta}
{\int^{+1}_{-1}\frac{d^2\mathcal{B}(B\to
K^{(*)}\ell^+\ell^-)}{d\hat{s}dcos\theta}dcos\theta}.
\end{eqnarray}
In the SM, the $\mathcal{A}_{FB}$ vanishes in $B\to K\ell^+\ell^-$
decays as shown by Eq.(\ref{BK}), since there is no term containing
$\hat{u}$ with an odd power. The RPV effect via the squark exchange
on $\mathcal{A}_{FB}(B\to K\ell^+\ell^-)$ also vanishes for the same
reason  as shown by Eq.(\ref{RPVBKmu}).

The total decay branching ratios of the pure leptonic $B_s$ decays
are calculated to be
\begin{eqnarray}
\mathcal{B}(B_s\to \ell^+\ell^-)&=&\mathcal{B}^{SM}(B_s\to
\ell^+\ell^-)\Bigg\{1+\frac{1}{|h_{SM}|^2}\bigg[2Re(h_{SM}\Lambda^*_{\tilde{u}})
+|\Lambda_{\tilde{u}}|^2
\bigg]\nonumber\\
&&+\frac{1}{|h_{SM}|^2}\bigg[Re(h_{SM}\Lambda^*_{\tilde{\nu}})\frac{1}{m_{\ell}}+
|\Lambda_{\tilde{\nu}}|^2
\bigg(\frac{1}{2m^2_{\ell}}-\frac{1}{m^2_{B_s}}\bigg)\bigg]\nonumber\\
&&+\frac{1}{|h_{SM}|^2}\bigg[-Re(h_{SM}\Lambda^{'*}_{\tilde{\nu}})\frac{1}{m_{\ell}}+
|\Lambda^{'}_{\tilde{\nu}}|^2
\bigg(\frac{1}{2m^2_{\ell}}-\frac{1}{m^2_{B_s}}\bigg)\bigg]\Bigg\}.
\end{eqnarray}
From this equation, we can see that $\mathcal{B}(B_s\to \ell^+\ell^-)$ could be enhanced
very much by the  s-channel RPV sneutrino exchange, but not by the  t-channel squark exchange.

\section{Numerical results and analysis}

Now we are ready to present our numerical results and analysis.
Firstly, we will show our estimations and compare them with the
relevant experimental data. Then, we will consider the RPV effects
to constrain the relevant RPV couplings from  the recent
experimental data. In addition, using the constrained parameter
spaces, we will give the RPV SUSY predictions for
$\mathcal{B}(B_s\to \ell^+\ell^-)$ and
$\mathcal{A}_{FB}(B\to K^{(*)}\ell^+\ell^-)$, which have not
been well measured yet.

For the form factors involving the $B\to K^{(*)}$ transitions, we
will use the recently  light-cone  QCD sum rules (LCSRs)  results \cite{BallZwicky},
 which are renewed with  radiative corrections to
the leading twist wave functions and SU(3) breaking effects.
 For the $q^2$ dependence of the form factors,
they can be parameterized in terms of simple formulae with two or
three parameters. The form factors $V, A_0$ and $T_1$
are  parameterized by
\begin{eqnarray}
F(\hat{s})=\frac{r_1}{1-\hat{s}/\hat{m}^2_{R}}+\frac{r_2}{1-\hat{s}/\hat{m}^2_{fit}}.\label{r12mRfit}
\end{eqnarray}
For the form factors $A_2, \tilde{T}_3, f_+$ and $f_T$, it is more
appropriate to expand to second order around the pole, yielding
\begin{eqnarray}
F(\hat{s})=\frac{r_1}{1-\hat{s}/\hat{m}^2}+\frac{r_2}{(1-\hat{s}/\hat{m})^2},\label{r12mfit}
\end{eqnarray}
where $\hat{m}=\hat{m}_{fit}$ for $A_2$ and $\tilde{T}_3$, and
$\hat{m}=\hat{m}_{R}$ for $f_+$ and $f_T$.  The fit formula
 for $A_1,T_2$ and $f_0$ is
\begin{eqnarray}
F(\hat{s})=\frac{r_2}{1-\hat{s}/\hat{m}^2_{fit}}.\label{r2mfit}
\end{eqnarray}
The form factor $T_3$ can be obtained by
$T_3(\hat{s})=\frac{1-\hat{m}_{K^*}}{\hat{s}}[\widetilde{T}_3(\hat{s})-T_2(\hat{s})]$.
All the corresponding  parameters for these form factors are
collected in Table I.  In the following numerical data analyses, the uncertainties
induced by $F(0)$\cite{BallZwicky} are also
considered.


\begin{table}[ht]
\centerline{\parbox{15cm}{\small Table I: Fit for form factors
involving the $B\to K^{(*)}$ transitions valid for general $q^2$
\cite{BallZwicky}.}} \vspace{0.3cm}
\begin{center}
\begin{tabular}{cccccccc}\hline\hline
$F(\hat{s})$&$~F(0)~$&$~\Delta_{tot}~$&$~r_1~$&$~m_R^2~$&$~r_2~$&$~m^2_{fit}~$&~fit
Eq.\\\hline $f_+^{B\to
K}$&$0.331$&$0.041$&$0.162$&$5.41^2$&$0.173$&&(\ref{r12mfit})\\\hline
$f_T^{B\to
K}$&$0.358$&$0.037$&$0.161$&$5.41^2$&$0.198$&&(\ref{r12mfit})\\\hline
$f_0^{B\to
K}$&$0.331$&$0.041$&&&$0.330$&$37.46$&(\ref{r2mfit})\\\hline
$V^{B\to
K^*}$&$0.411$&$0.033$&$0.923$&$5.32^2$&$-0.511$&$49.40$&(\ref{r12mRfit})\\\hline
$A_0^{B\to
K^*}$&$0.374$&$0.033$&$1.364$&$5.28^2$&$-0.990$&$36.78$&(\ref{r12mRfit})\\\hline
$A_1^{B\to
K^*}$&$0.292$&$0.028$&$$&$$&$0.290$&$40.38$&(\ref{r2mfit})\\\hline
$A_2^{B\to
K^*}$&$0.259$&$0.027$&$-0.084$&$$&$0.342$&$52.00$&(\ref{r12mfit})\\\hline
$T_1^{B\to
K^*}$&$0.333$&$0.028$&$0.823$&$5.32^2$&$-0.491$&$46.31$&(\ref{r12mRfit})\\\hline
$T_2^{B\to
K^*}$&$0.333$&$0.028$&$$&$$&$0.333$&$41.41$&(\ref{r2mfit})\\\hline
$\widetilde{T}_3^{B\to
K^*}$&$0.333$&$0.028$&$-0.036$&$$&$0.368$&$48.10$&(\ref{r12mfit})\\\hline
\end{tabular}
\end{center}
\end{table}

The other input parameters and the experimental data are collected
in Table II and III, respectively. In our numerical results,
if not specified, we will study physics observables in the region
$s>0.1GeV^2$,  and use the input parameters and the experimental
data which are varied randomly within $1\sigma$ and $2\sigma$
variance, respectively. We assume that only one sfermion contributes
at one time with a mass of 100 $GeV$. As for other values of the
sfermion masses, the bounds on the couplings in this paper can be
easily obtained by scaling them with factor $\tilde{f}^2\equiv
(\frac{m_{\tilde{f}}}{100GeV})^2$.

\begin{table}[h]
\centerline{\parbox{16.2cm}{\small Table II: Default values of the
input parameters and the $\pm1 \sigma$ error bars for the sensitive
parameters used in our numerical calculations.}} \vspace{0.3cm}
\begin{center}
\begin{tabular}{lc}\hline\hline
$m_{B_s}=5.370~GeV,~~m_{B_d}=5.279~GeV,~~m_{B_u}=5.279~GeV,~~m_W=80.425~GeV,$& \\
$m_{K^\pm}=0.494~GeV,~~m_{K^0}=0.498~GeV,~~m_{K^{*\pm}}=0.892~GeV,~~m_{K^{*0}}=0.896~GeV,$& \\
$\overline{m}_b(\overline{m}_b)=(4.20\pm0.07)~GeV,~~\overline{m}_s(2GeV)=(0.095\pm0.025)~GeV,$& \\
$\overline{m}_u(2GeV)=0.0015\sim
0.003~GeV,~\overline{m}_d(2GeV)=0.003\sim
0.007~GeV,$& \\
$m_e=0.511\times10^{-3}~GeV,~~m_\mu=0.106~GeV,$~~
$m_{t,pole}=174.2\pm3.3~GeV. $& \cite{PDG}\\ \hline
$\tau_{B_s}=(1.466\pm0.059)~ps,~~\tau_{B_{d}}=(1.530\pm
0.009)~ps,~~\tau_{B_{u}}=(1.638\pm 0.011)~ps.$& \cite{PDG}\\\hline
$|V_{tb}|\approx0.99910,~~|V_{ts}|=0.04161^{+0.00012}_{-0.00078}.$& \cite{PDG}\\
\hline $\mbox{sin}^2\theta_W=0.22306,~~\alpha_e=1/137.$&
\cite{PDG}\\\hline $f_{B_s}=0.230\pm0.030~GeV.$&
\cite{fBs}\\\hline\hline
\end{tabular}
\end{center}
\end{table}
\begin{table}[h]
\centerline{\parbox{12.3cm}{\small Table III: The SM predictions and the
experimental data  for  $B\to
K^{(*)}\ell^+\ell^-$ and $B_s\to \ell^+\ell^-$
\cite{ba1,be1,HFAG,PDG,cdf}.}} \vspace{0.4cm}
\begin{center}
\begin{tabular}
{l|c|c}\hline\hline & SM prediction value & Experimental
data\\\hline
 $\mathcal{B}(B \to K\mu^+\mu^-)$&0.610$\times10^{-6}$ &$(0.561^{+0.066}_{-0.061})\times10^{-6}$\\
$\mathcal{B}(B \to Ke^+e^-)$&$0.610\times10^{-6}$&$(0.380^{+0.073}_{-0.067})\times10^{-6}$\\
 $\mathcal{B}(B \to K^*\mu^+\mu^-)$&$ 1.27\times10^{-6}$&$(1.44\pm0.23)\times10^{-6}$\\
 $\mathcal{B}(B \to K^*e^+e^-)$& $1.29\times10^{-6}$&$(1.25\pm0.27)\times10^{-6}$\\
 $\mathcal{B}(B_s\to \mu^+\mu^-)$& $3.72\times10^{-9}$&$<1.0\times10^{-7}
 $~(95\%$~\emph{C.} \emph{L.}$) \\
 $\mathcal{B}(B_s \to e^+e^-)$&$8.70\times10^{-14}$&$<5.4\times10^{-5}
 $~(90\%$~\emph{C.} \emph{L.}$)\\ \hline\hline
\end{tabular}
\end{center}
\end{table}

The branching ratios in the SM estimated with  the central  values of the input
parameters are presented in Table III, and the relevant experimental data
 \cite{ba1,be1,HFAG,PDG,cdf} are listed for comparison. From Table III, we
can find  that the branching ratios of the semileptonic decays roughly
consistent with the SM predictions. So,  there are still windows, however limited,  for RPV contributions.

We now turn to the RPV effects.  There are six RPV coupling products
contributing to four $B\to K^{(*)}\ell^+\ell^-$ and two $B_s\to
\ell^+\ell^-$ decay modes. We use the experimental data of the
branching ratios listed  in Table III to constrain the relevant RPV
coupling products.

\begin{figure}[ht]
\begin{center}
\includegraphics[scale=1]{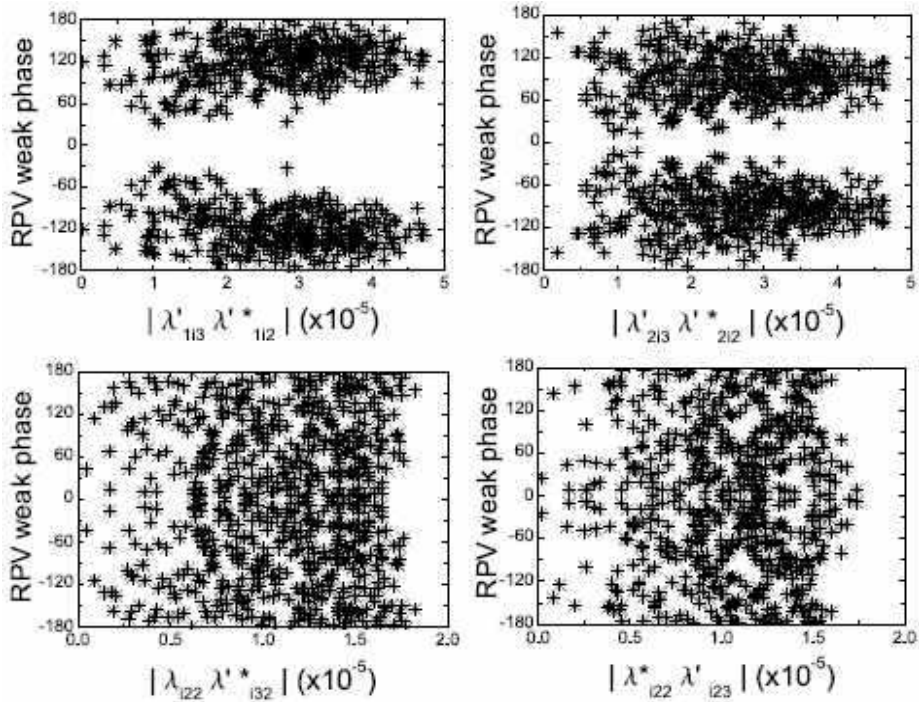}
\end{center}
\vspace{-0.6cm}
 \caption{ The allowed parameter spaces for the relevant
 RPV coupling products constrained by the measurements  listed in Table IV,
 and the RPV weak phase is given in degree.}
 \label{bounds}
\end{figure}

For the RPV couplings $\lambda_{i11}\lambda'^*_{i32}$ and
$\lambda^*_{i11}\lambda'_{i23}$, since their RPV weak phases are
found to have very small contributions to the physical observables,
we take their RPV weak phases to be free, and
 only give the upper limits for their modulus  which are listed in Table IV.
 In Fig.\ref{bounds}, we present our  bounds on the other four  RPV coupling  products.

From Fig.\ref{bounds}, we find that every RPV weak phase are not
constrained so much, but the modulus of the relevant RPV coupling
products are tightly upper limited.
 The upper limits for  the relevant RPV coupling products  by $B \to
K^{(*)}\ell^+\ell^-$ and $B _s\to \ell^+\ell^-$ decays are summarized  in
Table IV.   For comparison,  the existing bounds on these quadric
coupling products \cite{kundu,Allanach} are also listed.  From Table IV,  one can find that
our bounds are more restrict  than the previous ones.  It's  also noted  that the previous
bounds of $|\lambda_{i11}\lambda'^*_{i32}|$ and
$|\lambda^*_{i11}\lambda'_{i23}|$ are obtained  at the $M_{GUT}$
scale in the RPV mSUGRA model \cite{Allanach}.

\begin{table}[ht]
\centerline{\parbox{14.1cm}{Table IV: \small {Bounds for  the
relevant RPV coupling products by $B \to K^{(*)}\ell^+\ell^-$ and
$B_s \to \ell^+\ell^-$ decays for 100 $GeV$ sfermions, and previous
bounds are listed for comparison. }}} \vspace{0.5cm}
\begin{center}
\begin{tabular}{c|l|l}\hline\hline
Couplings&~~~~~~~~~Bounds [Processes]& Previous bounds [Processes]\\
\hline $|\lambda'_{1i3}\lambda'^*_{1i2}|$&$\leq4.7\times
10^{-5}~[B\to K^{(*)}e^+e^-]$ &${\leq 9.6\times 10^{-5}~[B\to K
e^+e^-}]$ \cite{kundu}\\
$|\lambda_{i11}\lambda'^*_{i32}| $&$\leq2.3\times 10^{-5}~[ B\to
K^{(*)}e^+e^-]$&$\leq1.5\times 10^{-3}$ \cite{Allanach}\\
$|\lambda^*_{i11}\lambda'_{i23}|$&$\leq2.3\times 10^{-5}~[B\to
K^{(*)}e^+e^-]$&$\leq1.9\times 10^{-4}$ \cite{Allanach}\\
$|\lambda'_{2i3}\lambda'^*_{2i2}| $&$\leq4.6\times 10^{-5}~_{[B\to
K^{(*)}\mu^+\mu^-]}^{[B_s \to \mu^+\mu^-]}$&$_{\leq9.7\times
10^{-5}~[B\to K\mu^+\mu^-]}^{\leq3.9\times
10^{-3}~[B_s\to \mu^+\mu^-]}$ \cite{kundu}\\
$|\lambda_{i22}\lambda'^*_{i32}| $&$\leq1.8\times 10^{-5}~_{[B\to
K^{(*)}\mu^+\mu^-]}^{[B_s \to \mu^+\mu^-]}$&$\leq2.7\times
10^{-5}~[B\to K\mu^+\mu^-]$ \cite{kundu}\\
$|\lambda^*_{i22}\lambda'_{i23}| $&$\leq1.7\times 10^{-5}~_{[B\to
K^{(*)}\mu^+\mu^-]}^{[B_s \to \mu^+\mu^-]}$&$\leq2.7\times
10^{-5}~[B\to K\mu^+\mu^-]$ \cite{kundu}\\\hline\hline
\end{tabular}
\end{center}
\end{table}

Using the constrained  parameter spaces shown in Table IV and
Fig.\ref{bounds}, we can predict the RPV effects on the other
quantities which have not been well measured yet in these processes.
We perform a scan over the input parameters and the new constrained
RPV coupling spaces to get the allowed ranges for
$\mathcal{B}(B_s\to\ell^+\ell^-)$ and
$\mathcal{A}_{FB}(B\to K^{(*)}\ell^+\ell^-)$  with
 the different RPV coupling products.  Our numerical results are
summarized  in Table V.


\begin{table}[ht]
\centerline{\parbox{16.6cm}{\small Table V: The theoretical
predictions for $\mathcal{B}(B_s\to \ell^+\ell^-)$ and
$\mathcal{A}_{FB}(B\to K^{(*)}\ell^+\ell^-)$ in the SM and the RPV
SUSY. The RPV SUSY predictions are obtained by the constrained
regions of the different RPV coupling products. The index  $g=1$ and
$2$ for $\ell=e$ and $\mu$, respectively.}} \vspace{0.3cm}
\begin{center}\small{
\begin{tabular}{l|c|c|c|c}\hline\hline
&SM value
&$\lambda'_{gi3}\lambda'^*_{gi2}$&$\lambda_{igg}\lambda'^*_{i32}$&$\lambda^*_{igg}\lambda'_{i23}$\\\hline
$\mathcal{B}(B_s\to
e^+e^-)$&$[5.9,11.2]\times10^{-14}$&$[5.3,17.0]\times 10^{-14}$
&$\leq2.2\times10^{-7}$&$\leq2.2\times10^{-7}$\\\hline
$\mathcal{A}_{FB}(B\to Ke^+e^-)$&$0$&$0$
&$[-4.1,5.0]\times10^{-5}$&$[-4.0,5.0]\times10^{-5}$\\\hline
$\mathcal{A}_{FB}(B\to
K^*e^+e^-)$&$[0.12,0.18]$&$[0.00,0.29]$ &$[0.12,0.18]$&$[0.13,0.18]$
\\\hline
$\mathcal{B}(B_s\to
\mu^+\mu^-)$&$[2.7,4.8]\times10^{-9}$&$[2.1,6.8]\times10^{-9}$
&$<1.0\times10^{-7}$&$<1.0\times10^{-7}$\\\hline
$\mathcal{A}_{FB}(B\to K\mu^+\mu^-)$&$0$&$0$
&$[-7.0,6.9]\times10^{-3}$&$[-6.8,7.0]\times10^{-3}$\\\hline
$\mathcal{A}_{FB}(B\to
K^*\mu^+\mu^-)$&$[0.13,0.18]$&$[0.02,0.76]$
&$[0.13,0.18]$&$[0.13,0.18]$
\\\hline \hline
\end{tabular}}
\end{center}
\end{table}

From the Table V, we can find some salient features of the numerical
results.

 \begin{itemize}
\item I.   As shown by Fig.\ref{purefig}(b), the contributions of
$\lambda'_{1i3}\lambda'^{*}_{1i2}$ and
$\lambda'_{2i3}\lambda'^{*}_{2i2}$ to $\mathcal{B}(B_s\to e^+e^-)$
and $\mathcal{B}(B_s\to \mu^+\mu^-)$, respectively,  arise from
t-channel  squark exchange . After Fierz transformation, the
effective Hamiltonian due to the  t-channel squark exchange is
proportional to
$\Big(\bar{s}\gamma^{\mu}P_Rb\Big)\Big(\bar{\ell}\gamma_{\mu}P_L\ell\Big)$,
which contribution to $\mathcal{B}(B_s\to \ell^+\ell^-)$ is
suppressed by $m^2_{\ell}/m^2_B$ due to helicity suppression.
Therefore $\mathcal{B}(B_s\to e^+e^-,\mu^+\mu^-)$ will not be
enhanced so much by the t-channel squark exchanging RPV
contributions. However, the effective Hamiltonian of s-channel
sneutrino exchange would be
$\Big(\bar{s}(1\pm\gamma_5)b\Big)\Big(\bar{\ell}(1\mp
\gamma_5)\ell\Big)$, which contributions are not suppressed by
$m^2_{\ell}/m^2_B$. So that, both  $\mathcal{B}(B_s\to e^+e^-)$ and
$\mathcal{B}(B_s\to \mu^+\mu^-)$ could be enhanced to order
$10^{-7}$ by
$\lambda_{i11}\lambda_{i32}'^{*}(\lambda^*_{i11}\lambda'_{i23})$ and
$\lambda_{i22}\lambda_{i32}'^{*}(\lambda^{*}_{i22}\lambda'_{i23})$,
respectively.

\item II.
  Both $\mathcal{A}_{FB}(B\to Ke^+e^-)$ and
$\mathcal{A}_{FB}(B\to K\mu^+\mu^-)$ are zero in the SM.  The
RPV contributions to the asymmetries due to squark exchange are
also zero, while the sneutino exchange RPV contributions are
too small to be accessible at LHC.

\item III.  It is interesting to note that the RPV squark exchange
contributions have significant impacts on
$\mathcal{A}_{FB}(B\to K^*e^+e^-)$ and
$\mathcal{A}_{FB}(B\to K^*\mu^+\mu^-)$.
However, the sneutrino
exchange have negligible  effects on $\mathcal{A}_{FB}(B\to
K^*e^+e^-)$ and $\mathcal{A}_{FB}(B\to K^*\mu^+\mu^-)$.
\end{itemize}

 Recently,   the \textit{BABAR}
Collaboration \cite{ba1} has measured
\begin{eqnarray}
&&\mathcal{A}_{FB}(B^+\to
K^+\ell^+\ell^-)_{(s>0.1GeV^2)}=0.15^{+0.21}_{-0.23}\pm0.08,\\
&&\mathcal{A}_{FB}(B\to
K^*\ell^+\ell^-)_{(s>0.1GeV^2)}\geq0.55~~ (95\% C.L.),
\end{eqnarray}
and Belle Collaboration has measured   the integrated forward-backward
asymmetries $\widetilde{\mathcal{A}}_{FB}$ \cite{be2}
\begin{eqnarray}
&&\widetilde{\mathcal{A}}_{FB}(B^+\to
K^+\ell^+\ell^-)=0.10\pm0.14\pm0.01,\\
&&\widetilde{\mathcal{A}}_{FB}(B\to
K^*\ell^+\ell^-)=0.50\pm0.15\pm0.02,
\end{eqnarray}
where $\widetilde{\mathcal{A}}_{FB}$ is slightly different from
$\mathcal{A}_{FB}$, and defined by
\begin{eqnarray}
\widetilde{\mathcal{A}}_{FB}(B\to
K^{(*)}\ell^+\ell^-)=\frac{\int \frac{d^2\mathcal{B}(B\to
K^{(*)}\ell^+\ell^-)}{dcos\theta d\hat{s} }sign(cos\theta)dcos\theta
d\hat{s}}{\int \frac{d^2\mathcal{B}(B\to
K^{(*)}\ell^+\ell^-)}{dcos\theta d\hat{s} }dcos\theta d\hat{s}}.
\end{eqnarray}
We find that the SM predictions for $ \mathcal{A}_{FB}(B\to
K^*\ell^+\ell^-)$  are consistent  with  both the measurements
within error bars.

\begin{figure}[h]
\begin{center}
\includegraphics[scale=1]{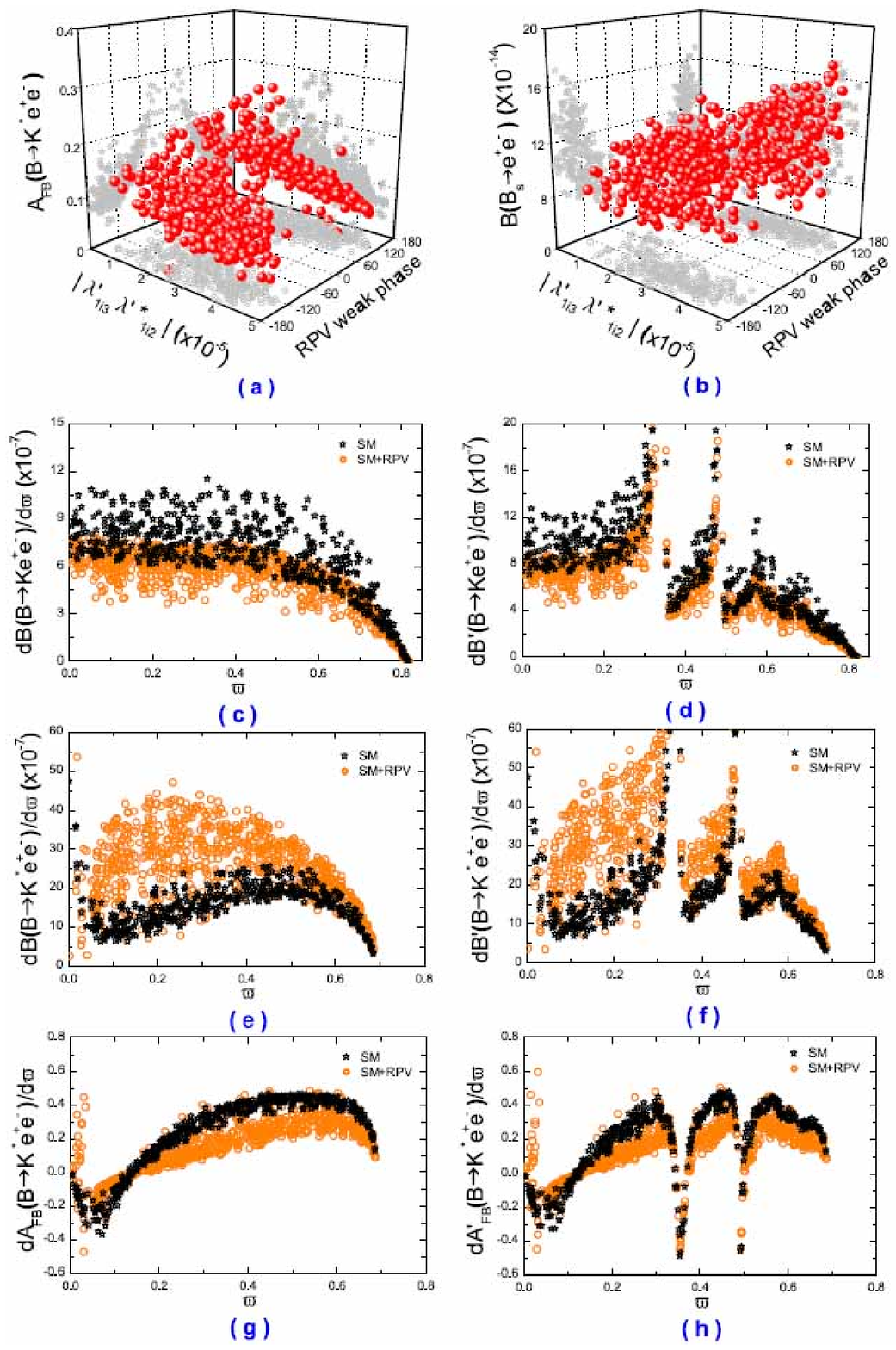}
\end{center}
\caption{ The effects of RPV coupling $\lambda'_{1i3}
 \lambda'^*_{1i2}$ due to the squark exchange
 in $B\to K^{(*)}e^+e^-$ and $B_s\to e^+e^-$ decays.
 The primed observables are given with $\psi(nS)$
 VMD contributions, and $\varpi$
 denote $\hat{s}$.} \label{figelplps}
\end{figure}

\begin{figure}[h]
\begin{center}
\includegraphics[scale=1]{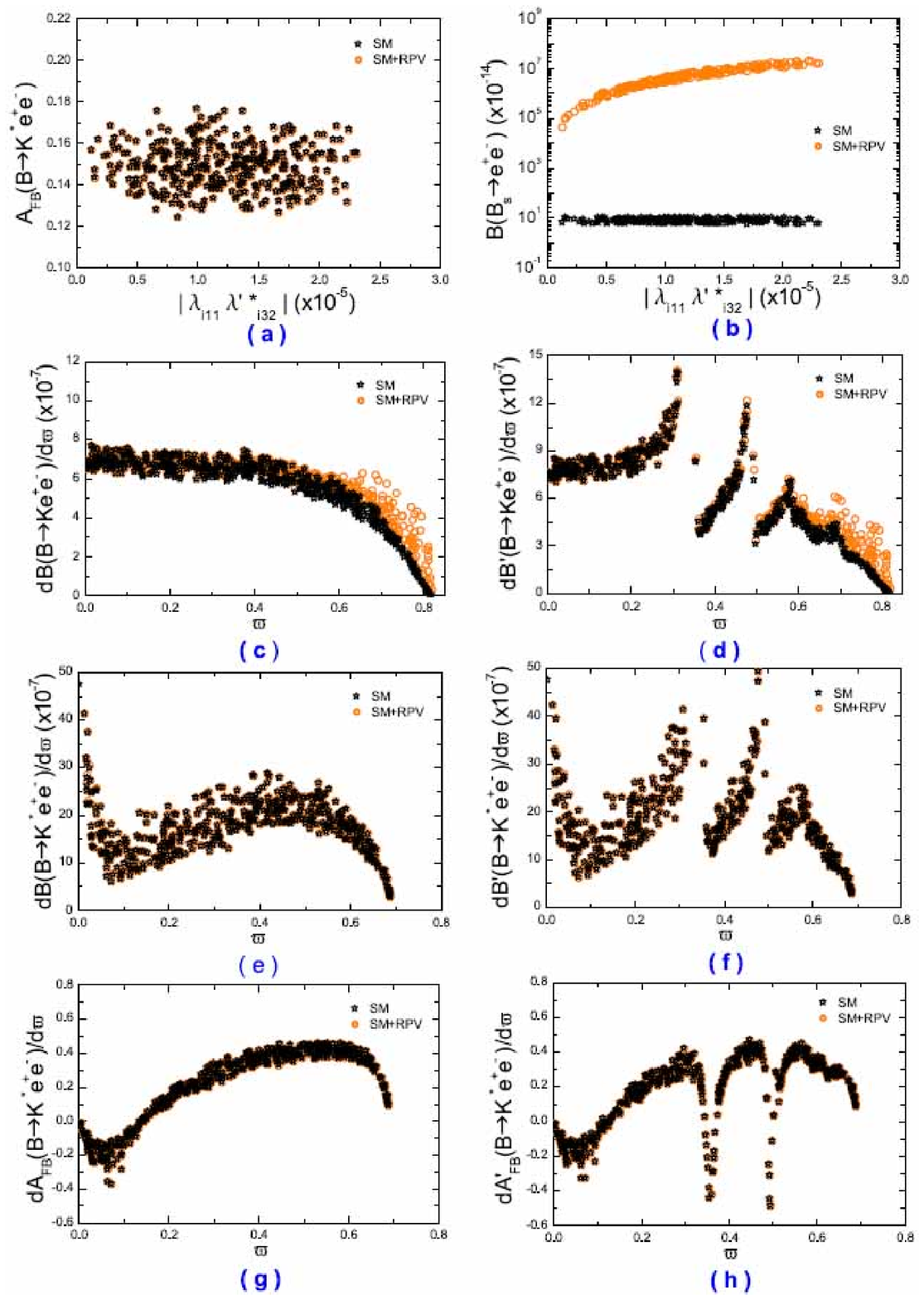}
\end{center}
\caption{ The effects of RPV coupling $\lambda_{i11}
 \lambda'^*_{i32}$ due to the sneutrino exchange in
 $B\to K^{(*)}e^+e^-$ and $B_s\to e^+e^-$ decays.
 }\label{figellps}
\end{figure}

\begin{figure}[h]
\begin{center}
\includegraphics[scale=1]{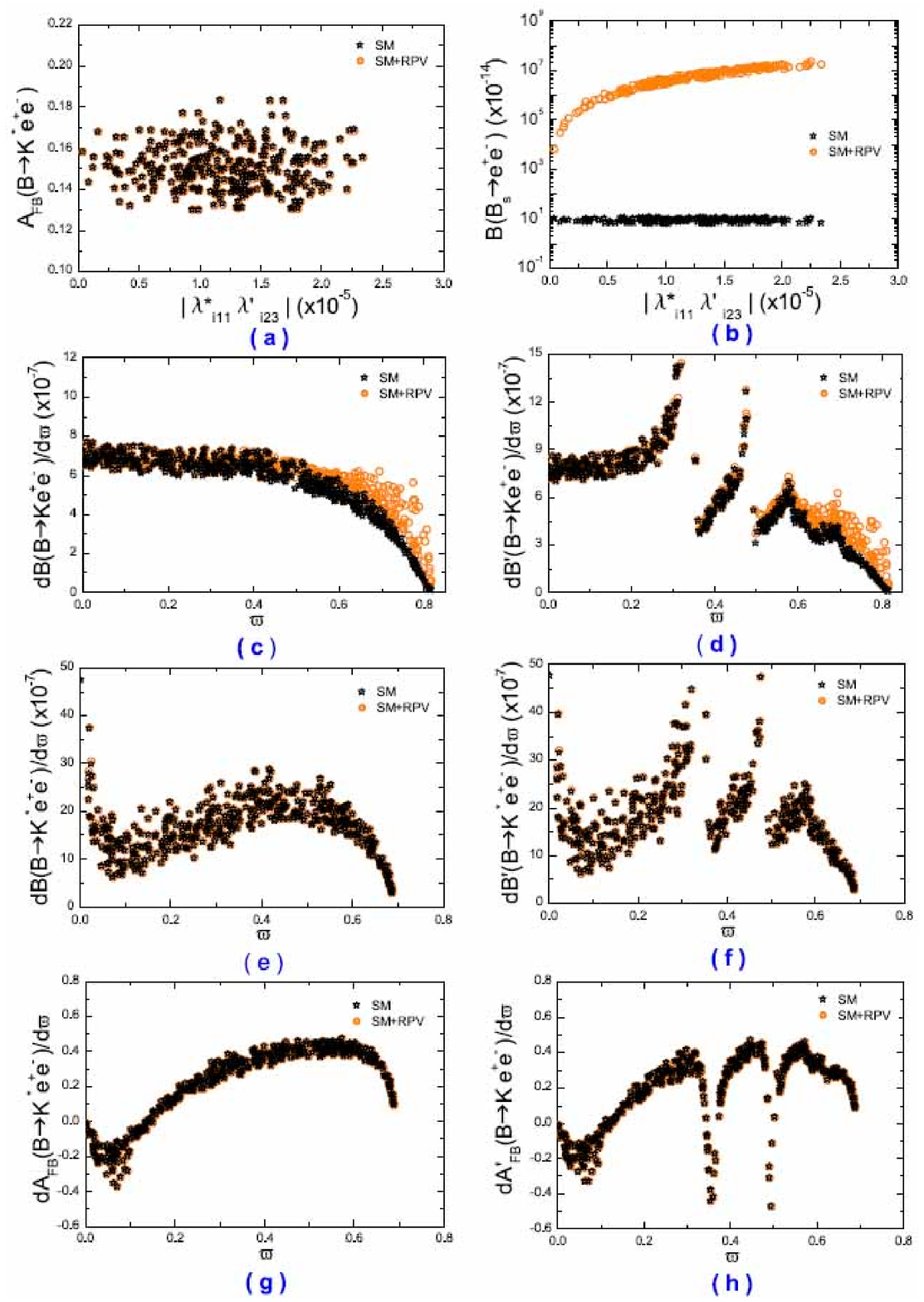}
\end{center}
\caption{ The effects of RPV coupling $\lambda^*_{i11}
 \lambda'_{i23}$ due to the sneutrino exchange in
 $B\to K^{(*)}e^+e^-$ and $B_s\to e^+e^-$ decays. }
\label{figelslp}
\end{figure}

\begin{figure}[h]
\begin{center}
\includegraphics[scale=1]{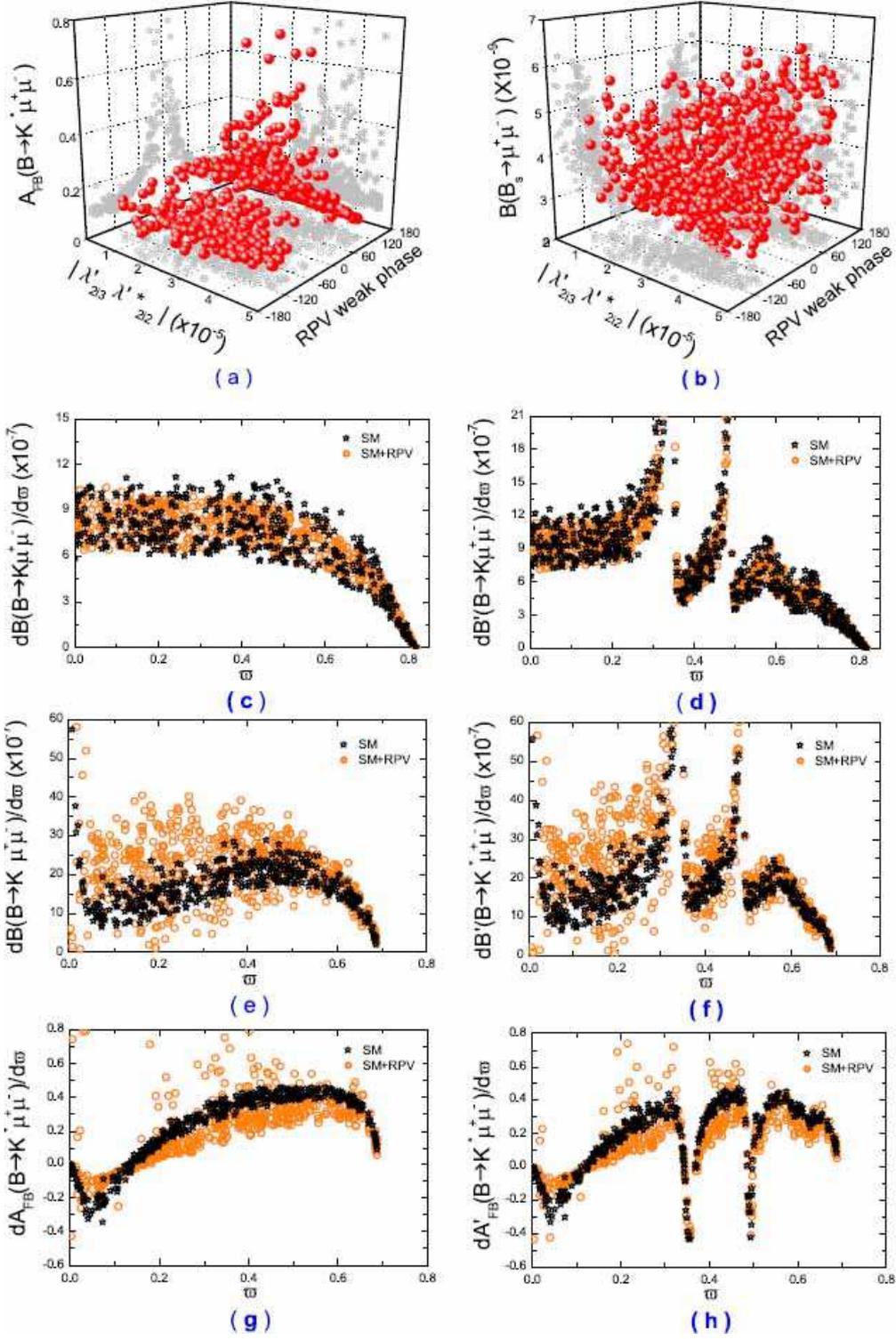}
\end{center}
\caption{ The effects of RPV coupling $\lambda'_{2i3}
 \lambda'^*_{2i2}$ due to the squark exchange in $B\to K^{(*)}\mu^+\mu^-$
 and $B_s\to \mu^+\mu^-$ decays.}\label{figulplps}
\end{figure}

\begin{figure}[h]
\begin{center}
\includegraphics[scale=1]{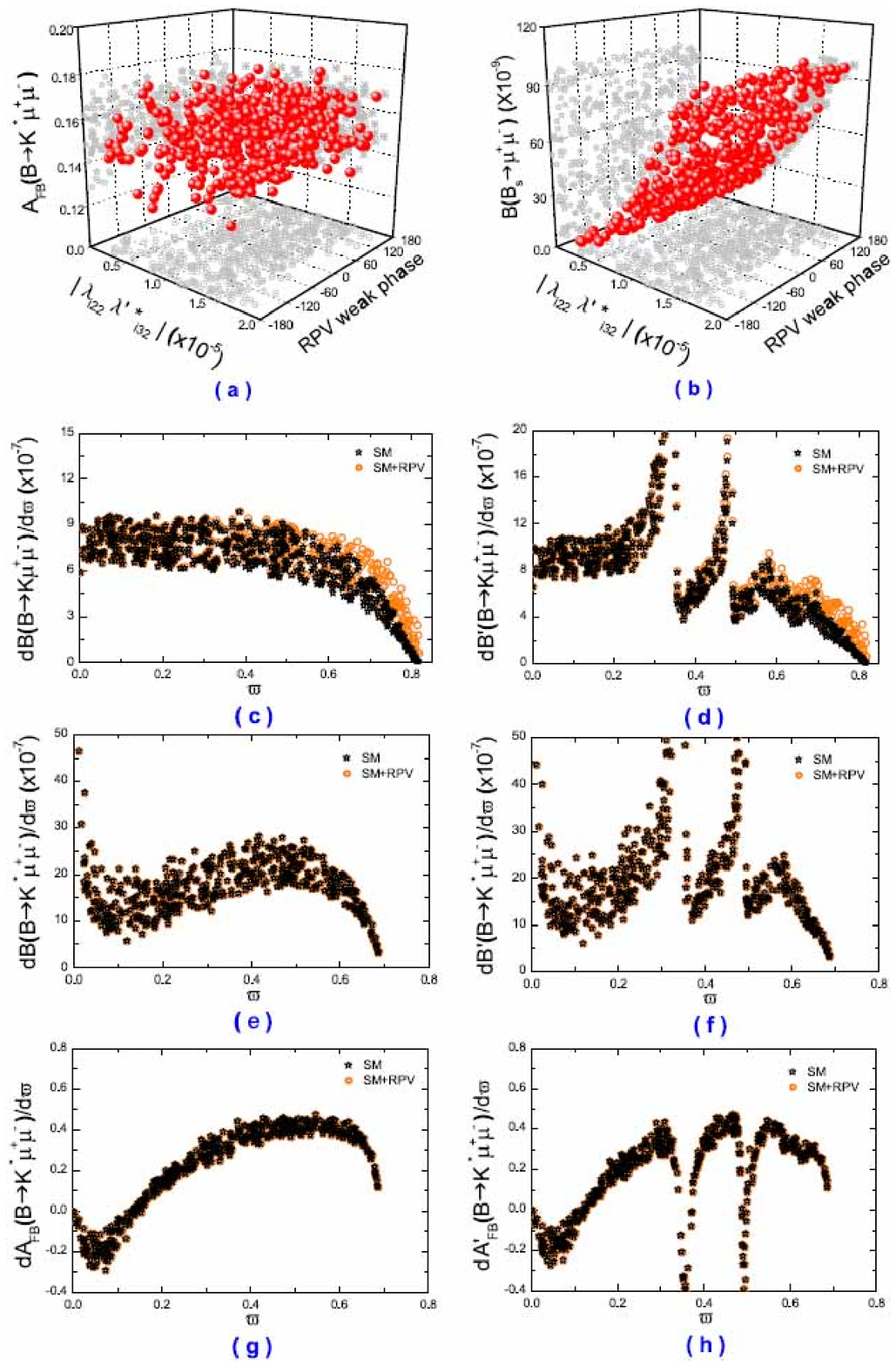}
\end{center}
\caption{ The effects of RPV coupling $\lambda_{i22}
 \lambda'^*_{i32}$ due to the sneutrino exchange in $B\to K^{(*)}\mu^+\mu^-$
 and $B_s\to \mu^+\mu^-$ decays.
}\label{figullps}
\end{figure}

\begin{figure}[h]
\begin{center}
\includegraphics[scale=1]{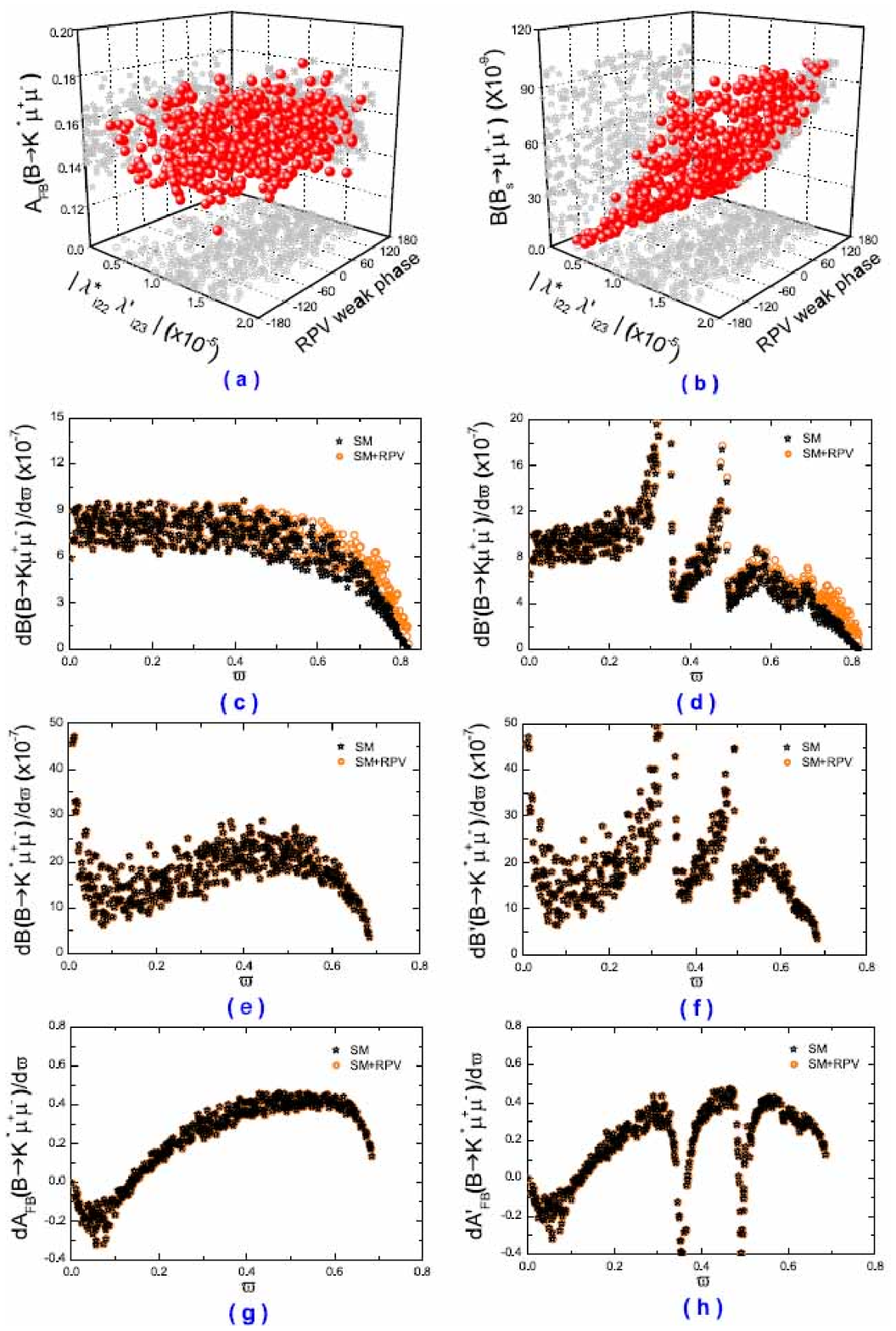}
\end{center}
\caption{The effects of RPV coupling $\lambda^*_{i22}
 \lambda'_{i23}$ due to the sneutrino exchange in $B\to K^{(*)}\mu^+\mu^-$
 and $B_s\to \mu^+\mu^-$ decays.}\label{figulslp}
\end{figure}

In Figs.\ref{figelplps}-\ref{figulslp}, we present correlations
between the physical observable $\mathcal{B}$, $\mathcal{A}_{FB}$
and the parameter spaces of the different RPV coupling products by
the three-dimensional scatter plots and the two-dimensional scatter
plots, respectively. The dilepton invariant mass distribution and
the normalized forward-backward asymmetry are given with VMD
contribution excluded in terms of $d\mathcal{B}/d\hat{s}$ and
$d\mathcal{A}_{FB}/d\hat{s}$,  and
 included in  $d\mathcal{B}'/d\hat{s}$ and
 $d\mathcal{A}'_{FB}/d\hat{s}$, respectively. From
Figs.\ref{figelplps}-\ref{figulslp}, one can find  the correlations
of these physical observables with RPV coupling products. In
Figs.\ref{figellps}-\ref{figelslp}, since the influences of the RPV
weak phases are very small, we take the RPV weak phases randomly
varied in $[-\pi,\pi]$ and only give the change trends of the
physical observables with the relative modulus in the
two-dimensional scatter plots.

At first, we will discuss  plots of  Fig.\ref{figelplps} in detail.
The three-dimensional scatter plot  Fig.\ref{figelplps}(a) shows
$\mathcal{A}_{FB}(B\to K^{*}e^+e^-)$ correlated with
$|\lambda'_{1i3}\lambda'^*_{1i2}|$ and its phase
$\phi_{\spur{R_p}}$.  We also give projections on three vertical
 planes, where the $|\lambda'_{1i3}\lambda'^*_{1i2}|$-$\phi_{\spur{R_p}}$ plane displays the
 constrained
 regions of $\lambda'_{1i3} \lambda'^*_{1i2}$ as the first plot of Fig.\ref{bounds}.
 It's shown that $\mathcal{A}_{FB}(B\to K^*e^+e^-)$ is decreasing
 with $|\lambda'_{1i3}\lambda'^*_{1i2}|$
 on the $\mathcal{A}_{FB}(B\to K^{*}e^+e^-)$-$|\lambda'_{1i3} \lambda'^*_{1i2}|$ plane.
 From the $\mathcal{A}_{FB}(B\to
 K^{*}e^+e^-)$-$\phi_{\spur{R_p}}$ plane,
 we can see that $\mathcal{A}_{FB}(B\to K^*e^+e^-)$ is decreasing with  $|\phi_{\spur{R_p}}|$.
  The recent measurement of $\mathcal{A}_{FB}(B\to K^*\ell^+\ell^-)$
   favors $|\lambda'_{1i3}\lambda'^*_{1i2}|$$\sim(2\sim4)\times10^{-5}$ with
   small $\phi_{\spur{R_p}}$. However $B_s\to e^+e^-$ is
   remained inaccessible  at LHC. From plots Fig.\ref{figelplps}(c-h), we can
   find that the theoretical uncertainties in the calculation of
   the SM contributions are still very large, nevertheless, for $d\mathcal{B}(B\to
   K^{(*)}e^+e^-)/d\hat{s}$ and $d\mathcal{A}_{FB}(B\to
   K^*e^+e^-)/d\hat{s}$, the $\lambda'_{1i3}\lambda'^{*}_{1i2}$
   contributions are  distinguishable from the hadronic uncertainties.

 Fig.\ref{figellps} and Fig.\ref{figelslp} are the contributions of
 $\lambda_{i11}\lambda'^{*}_{i32}$ and
 $\lambda^*_{i11}\lambda'_{i23}$, respectively. It is
 interesting to note that the t-channel RPV sneutrino exchange could
 enhance $\mathcal{B}(B_s\to e^+e^-)$ by about 6 orders of magnitude \Big(Fig.\ref{figellps}(b) and
 Fig.\ref{figelslp}(b)\Big). Moreover, they could give distinguishable
 contributions to $d\mathcal{B}(B\to
   Ke^+e^-)/d\hat{s}$ in the high $\hat{s}$ region.  For other observables, their
   contributions are indistinguishable from hadronic
   uncertainties.

The RPV squark exchange contributions to $B\to
K^{(*)}\mu^+\mu^-$ and $B_s\to\mu^+\mu^-$ are presented in
Fig.\ref{figulplps}. Such contributions could give large
$\mathcal{A}_{FB}(B\to K^*\mu^+\mu^-)$ in favor of Belle and
\textit{BABAR} measurements, however, small corrections to
$\mathcal{B}(B_s\to \mu^+\mu^-)$.

  The RPV sneutrino   exchange contributions to $B\to
  K^*\mu^+\mu^-$ and $B_s\to\mu^+\mu^-$ are displayed in
  Fig.\ref{figullps} and Fig.\ref{figulslp}, respectively.  From plots in these two figures,
  we find $\mathcal{B}(B_s\to\mu^+\mu^-)$
  is very sensitive  to such contributions. Within the constrained
  parameter space for $\lambda_{i22}\lambda'^{*}_{i32}$ and
  $\lambda^*_{i22}\lambda'_{i23}$, RPV contributions can
  enhance $\mathcal{B}(B_s\to\mu^+\mu^-)$ by two orders,
  which could be accessible at LHC and Tevatron in the forthcoming
  years. We also note that these contributions to  other
  observables are small, regarding to the large theoretical uncertainties.

  It is worth to note that,  in the BABAR measurement\cite{ba1},  the value of $\mathcal{A}_{FB}(B\to
K^*\ell^+\ell^-)$ with $ 95 \%~\emph{C.L.}$ lower limit is slightly
above the SM prediction in low $\hat{s}$ region. We have found that  the RPV couplings
$\lambda'_{1i3}\lambda'^{*}_{1i2}$ and
$\lambda'_{2i3}\lambda'^{*}_{2i2}$ could enhance
$\mathcal{A}_{FB}(B\to K^*\ell^+\ell^-)$ to  accommodate  the possible discrepancy, which
are shown by  the following Fig.4(g) and Fig.7(g), respectively.

 Generally, the predictions of $B$ decays suffer from  many theoretical
 uncertainties.   Recently,   beyond naive factorization,  $B\to K^{\ast} $ 
 have been investigated  with the QCD factorization\cite{QCDF} framework 
 in  \cite{Feldm1,   Feldm2},  and then studied with Soft Collinear Effective Theory (SCET) \cite{SCET} 
 in Ref.\cite{AliZhu}.  However, the power corrections are found to be associated with  large theoretical uncertainties,  and we are conservative to the naive factorization in this paper.
 
To probe new physics effects, it would be very useful to measure correlative observables, for
example, $d\mathcal{A}_{FB}(B\to K^*\mu^+\mu^-)/d\hat{s}$,
$d\mathcal{B}(B\to K^*\mu^+\mu^-)/d\hat{s}$ and
$\mathcal{B}(B_s\to\mu^+\mu^-)$, since correlations  among
these observables could provide very strict bound on new physics
models. At present, one may have to wait for the error bars  in the
measurements of these observables to come down and more channels to
be measured. With the operation of $B$ factory experiments, large
amounts of experimental data on hadronic $B$ meson decays are being
collected, and measurements of previously known observable will
become more precise. From the comparison of our predictions in
Figs.\ref{figelplps}-\ref{figulslp} with the near future
experiments,   one will obtain more stringent bounds on the products
of RPV couplings. On the other hand, the RPV SUSY predictions of
other decays will become more precise by the more stringent bounds
on the RPV couplings.

\section{Conclusions}
In conclusion, the decays $B \to K^{(*)}\ell^+\ell^-$ and $B_s\to
\ell^+\ell^-$  are very promising means to probe effects of new
physics scenarios. In this paper, we have studied the $B \to
K^{(*)}\ell^+\ell^-$ and $B_s\to \ell^+\ell^-$ decays in  the RPV
SUSY model. We have obtained fairly constrained parameter spaces of
the RPV coupling products from the present experimental data of $B
\to K^{(*)}\ell^+\ell^-$ and $B_s\to \ell^+\ell^-$ decays, and found
these constraints are stronger than the existing ones, which may be
useful for further studies of the RPV SUSY phenomenology.
Furthermore, using the constrained parameter spaces, we have
presented the RPV effects on the forward-backward asymmetries of
$B\to K^{(*)}\ell^+\ell^-$ and the branching ratios of the pure
leptonic $B_s$ decays. Our results of $\mathcal{A}_{FB}(B \to
K^{(*)}\ell^+\ell^-)$ agree with the recent experimental data. It is
shown that $\mathcal{B}(B_s\to \ell^+\ell^-)$ could be enhanced
several orders by the RPV couplings from the sneutrino exchange.
Since we have poorly experimental information about the pure
leptonic decay $B_s\to \ell^+\ell^-$, further refined measurements
of $\mathcal{B}(B_s\to \ell^+\ell^-)$
 can further restrict the constrained space of the four RPV couplings
 from the sneutrino exchange.
We have also compared the SM predictions with the RPV  predictions
about dilepton invariant mass spectra and the normalized
forward-backward asymmetries in $B \to K^{(*)}\ell^+\ell^-$ decays.
We have found that $d\mathcal{B}(B\to Ke^+e^-)/d\hat{s}$
could be mildly decreased by the RPV coupling
$\lambda'_{1i3}\lambda'^{*}_{1i2}$, and the
$\lambda'_{2i3}\lambda'^{*}_{2i2}$ contributions to
$d\mathcal{B}(B\to K\mu^+\mu^-)/d\hat{s}$ are
indistinguishable from the hadronic uncertainties. The other four
RPV couplings due to the sneutrino exchange have distinguishable
effects on $d\mathcal{B}(B\to K\ell^+\ell^-)/d\hat{s}$ at
high $\hat{s}$. For the $B\to K^*\ell^+\ell^-$ decays, the
squark exchange contributions have distinguishable effects on
$d\mathcal{B}/d\hat{s}$ and $d\mathcal{A}_{FB}/d\hat{s}$, but the
contributions from the sneutrino exchange have no apparent effects
on them.  The results in this paper could be useful for probing RPV SUSY effects
and will  correlate strongly  with searches for direct RPV signals at LHC in the forthcoming year.

\section*{Acknowledgments}
The work is supported  by National Science Foundation under contracts
No.10305003 and No.10675039, and the NCET Program sponsored by
Ministry of Education, China, under NCET-04-0656.



\end{document}